%% file: paper.tex
\documentclass[a4paper]{llncs}
\usepackage{amsmath}

\usepackage{enumitem}

\usepackage[nounderscore]{syntax}
\usepackage[caption=false]{subfig}

\usepackage{multirow}

\usepackage{algorithm}
\usepackage{algorithmicx}
\usepackage[noend]{algpseudocode}
\algblockdefx[Case]{Case}{EndCase}[1][]{\textbf{case} #1 \textbf{of}}{\textbf{end case}}
\algloopdefx[CaseItem]{CaseItem}
[1][default]{\texttt{#1:} }
\usepackage{mypackage}
\usepackage{chngcntr}
\usepackage{tcolorbox}

\usepackage{wrapfig}

\usepackage[svgnames]{xcolor}
\usepackage{colortbl}
\usepackage{pgf}
\usepackage{tikz}
\usetikzlibrary{arrows,shapes,fit}

\usepackage{todonotes}

\usepackage{url}

\newcommand{\indalt}[1][2]{\\\hspace*{#1em}\textbar\quad}
\newcommand{\ind}[1][2]{\\\hspace*{#1em}\quad}

\begin{document}
	
\counterwithout{lstlisting}{section}

\mainmatter

%\title{Dynamic Binding of Untrusted Parties on Blockchain: Executing Collaborative Business Processes Controlled by a Declarative DSL}
\title{Dynamic Role Binding in Blockchain-Based Collaborative Business Processes}

\author{\footnotesize Orlenys L\'opez-Pintado\inst{1}
\hspace*{-1ex}   \and Marlon Dumas\inst{1}
\hspace*{-1ex}   \and Luciano Garc\'ia-Ba\~nuelos\inst{1}
\hspace*{-1ex}   \and Ingo Weber\inst{2}}
\authorrunning{L\'opez-Pintado et al.}

\institute{University of Tartu, Estonia\\
\email{\{orlenyslp, marlon.dumas, luciano.garcia\}@ut.ee }
    \and
Data61, CSIRO, Sydney, Australia\\
  \email{ingo.weber@data61.csiro.au}
}

\maketitle

\begin{abstract}
Blockchain technology enables the execution of collaborative business processes involving mutually untrusted parties. Existing platforms allow such processes to be modeled using high-level notations and compiled into smart contracts that can be deployed on blockchain platforms. However, these platforms brush aside the question of who is allowed to execute which tasks in the process, either by deferring the question altogether or by adopting a static approach where all actors are bound to roles upon process instantiation. Yet, a key advantage of blockchains is their ability to support dynamic sets of actors. This paper presents a model for dynamic binding of actors to roles in collaborative processes and an associated binding policy specification language. The proposed language is  endowed with a Petri net semantics, thus enabling policy consistency verification. The paper also outlines an approach to compile policy specifications into smart contracts for enforcement. An experimental evaluation shows that the cost of policy enforcement increases linearly with the number of roles and constraints.

\end{abstract}

\input{introduction}

\input{background}

\input{model}

\input{consistency}
\input{implementation}

\input{conclusion}

\bibliographystyle{spmpsci}
\bibliography{bibliography}

\end{document}

%% file: introduction.tex
% !TEX root = ../paper.tex

\section{Introduction}
\label{sec:introduction}

Access control is an essential aspect in the design and execution of business processes. Mainstream Business Process Management Systems (BPMSs) rely on static Role-Based Access Control (RBAC) models. In these models, any worker who plays a role is allowed to perform any task associated to this role in any instance of the process, modulo additional constraints such as separation of duties~\cite{RussellAHE05}. 
This approach is unsuitable for collaborative inter-organizational processes involving untrusted actors. For example, a buyer may trust a given carrier but not others, even though they all play the same role.
%In this context, classical RBAC models that rely on a trusted central authority (typically the employer of all users) are not suitable. 
%On the other hand, traditional schemes for dynamic allocation and delegation of tasks among actors  also fails as the trust cannot be delegated. 
%Additionally, some Business Process Management Systems (BPMSs) allow defining constraints such as separation of duties, or retaining chosen users~\cite{RussellAHE05}, e.g., to restrict/favor whether one user can perform groups of tasks.

Blockchain technology enables the execution of collaborative business processes involving untrusted actors~\cite{Mendling18}. Existing tools such as Caterpillar \cite{Lopez-PintadoGDWP18} and Lorikeet \cite{TranLW18}, support the definition of collaborative processes using high-level notations and their execution on top of blockchain platforms. 
However, these tools either do not support access control or they adopt a static role binding approach wherein all actors are bound to roles upon process instantiation.

%Blockchain technology emerges as a solution to solve the lack of trust in the collaborative processes~\cite{Mendling18}. Specifically, the process execution logic can be encoded as autonomous programs called smart contracts. Several existing blockchain-based systems such as Caterpillar \cite{Lopez-PintadoGD17} and Lorikeet \cite{TranLW18} address the issue with trust by executing collaborative processes in the blockchain. 

%On the other hand, several models for role-based workflow binding have been proposed, but they do not address the blockchain specificities, e.g., the absence of trust, which requires explicit endorsements. 

\begin{wrapfigure}{r}{0.46\textwidth}
 	\vspace*{-4mm}
	\centering
	\includegraphics[width=0.44\columnwidth]{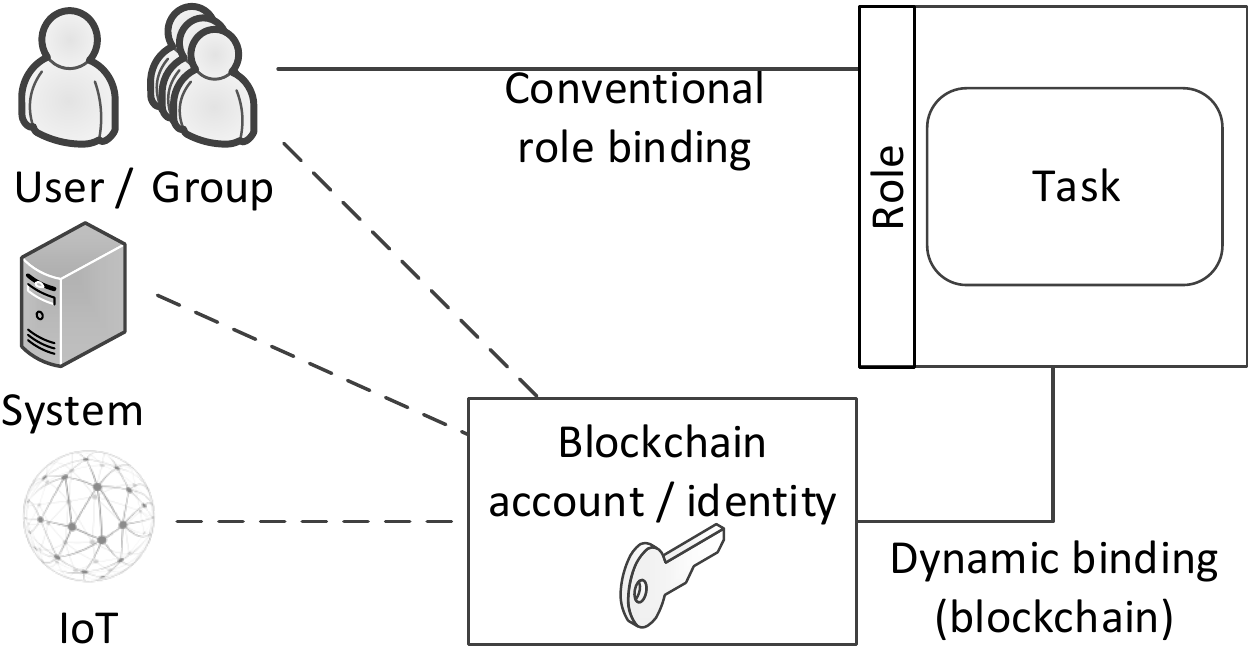}
	\caption{Relations between tasks, roles, blockchain accounts, and actors (blockchain case vs. conventional case).}
	\label{fig:concept-relationship}%
	\vspace*{-6mm}
\end{wrapfigure}

The characteristics of blockchain technology shift the role binding problem in two ways. First, rather than groups or individual users being bound to roles, we need to bind blockchain accounts (or identities) to roles,  as shown in Fig.~\ref{fig:concept-relationship}. These accounts, in turn, are controlled by users, groups, systems, or (IoT) devices. %This change renders previous solutions insufficient. 
Second and more significantly, in open blockchain networks, instances of a collaborative process are created by different actors, and each of these actors trusts one subset of actors but not others. Moreover, the set of actors changes dynamically and so do the trust relations. For example, a buyer may initially trust a carrier and agree to its appointment together with the supplier. But later, the buyer may lose this trust (e.g.\ if the carrier misses a deadline). Thereafter, the buyer may wish to re-bind the transportation task to another carrier, but this re-binding must be endorsed by the supplier. This example illustrates the need to support \emph{dynamic binding and un-binding of actors to roles} and \emph{collaborative binding of actors to roles} (buyer and supplier both need to agree on the carrier).

This paper proposes a role binding model and a binding policy specification language designed to support collaborative business processes in such open and untrusted environments, as well as an approach to compile policies into executable code. The semantics of the policy specification language is defined via a mapping to Petri nets, which enables the static verification of policies prior to their compilation. The proposed method has been implemented in Caterpillar \cite{Lopez-PintadoGDWP18} -- a blockchain-based execution engine that supports the Business Process Model and Notation (BPMN). The paper reports on an experimental evaluation aimed at assessing the cost of policy enforcement on the Ethereum blockchain.
%into smart contracts. 
% we empirically evaluate the cost to deploy and execute the smart contracts on the Ethereum blockchain for more than 200 policies.

The rest of the paper is structured as follows. Section \ref{sect:background} discusses basic concepts of blockchain technology and the limitations of existing role binding models for collaborative processes. Section~\ref{sect:model} describes the role binding model and policy language. Section~\ref{sect:policy} presents the semantics of the policy language and the policy verification approach. Finally, Section \ref{sect:implementation} discusses the implementation and evaluation, while  Section \ref{sect:conclusion} draws conclusions and sketches future work.

%% file: background.tex
% !TEX root = ../paper.tex

\section{Background and Related Work}
\label{sect:background}

\subsection{Blockchain Technology and Collaborative Processes}
\label{ssect:blockchain-bg}

A blockchain is a distributed append-only store of transactions distributed across computational nodes and structured as a linked list of blocks, each containing a set of transactions~\cite{2019-Blockchain-Book}.
A blockchain network is made up of nodes, a subset of which holds a replica of the data structure. %, and in a public permissionless blockchain network the nodes can join and leave as they please.
Clients use a blockchain system (a concrete network) by reading data from and submitting transactions to it. Submitted transactions are grouped into blocks, which are broadcast across the network to be appended to the blockchain.
%To be accepted, a transaction must be properly formed and signed by their creator. 
%No trust in individual clients or nodes is required, as the transactions are cryptographically signed and validated, and broadcast widely.
A consensus mechanism ensures tamper-proofness without assuming mutual trust between participants.
%Cryptographic methods and economic incentives 
 
 A smart contract is a program deployed on the blockchain, which may be invoked via a transaction~\cite{2019-Blockchain-Book}.
%smart contract is a computer program executed on the blockchain network by all the nodes. 
%For example, each transaction in the Ethereum blockchain is described by one or several smart contracts. %NOT TRUE btw.
In Ethereum, smart contracts are written in the Solidity language, which is compiled into bytecode and executed on the Ethereum Virtual Machine. 
The computational and data storage consumption of a transaction are measured in \emph{gas}, which translates to monetary costs for the transaction's sender. 
Each block has a \emph{gas limit} and hence gas directly impacts throughput.
%Contracts are deployed through transactions. %also costs gas. %, which is proportional from its bytecode size, is measured in gas. 
%During deployment, a smart contract is assigned a unique address, used by client applications to call its functions. 
%%Such execution also generates a transaction whose cost in gas, depends on the number and type of the operations performed. 
%External actors must hold an Ethereum account to deploy and interact with smart contracts. An account comprises a public address, i.e., a sort of user ID, which doubles as public key, and a private key to sign the transactions~\cite{wood2014ethereum}.
 
\enlargethispage{0.6\baselineskip}
Existing blockchain-based process management tools support the specification of collaborative processes using BPMN~\cite{Lopez-PintadoGDWP18,TranLW18} or domain-specific languages \cite{DBLP:FrantzN16}, and their execution via smart contracts. These systems focus mainly on the control-flow perspective. Lorikeet~\cite{TranLW18} %implemented from \cite{WeberXRGPM16}, 
implements a static access control mechanism, where roles are bound to accounts upon process instantiation. 
%The Caterpillar prototype \cite{Lopez-PintadoGDWP18} uses worklists to execute user tasks but lacks an access control mechanism. 
A method proposed in~\cite{Prybila0HW17} allows dynamic handoffs of process instances between actors, but does not support the specification and enforcement of permitted handoffs.
 %changes at runtime but assumes a participant monitors the execution and react in front of undesired changes. 
%Although not for process collaborations, some works rely on blockchain as a source of trustworthy data and with access control to its manipulation.
 %to protect data and implement access control systems. %WeberXRGPM16,
%However, dynamic bindings, specifically to handle the execution of collaborative processes in blockchain is still a problem to be addressed. 

\vspace{-0.9em} 
\subsection{Binding and Delegation Models for Collaborative Processes}
\vspace{-0.4em} 

%The problem of dynamic role binding and delegation  the process execution was widely addressed in the decade of the 2000s. However, challenges like the lack of trust and limited visibility stopped the organizations to fully adopt such techniques on the execution of inter-organizational processes. A recent survey \cite{Pourmirza17} shows that only 30\% out of BPMSs analyzed have addressed the concept of inter-organizational process. The authors point out that one of the main challenges is that the autonomy of organizations clashes with putting more trust in inter-organizational processes. Besides, further investigation is needed on the dynamism and flexibility in collaborative processes to satisfy the demands of modern-day business \cite{Grefen16}.

% architectures support bindings as abstraction to separate implementation from high-level descriptions
The question of dynamic role binding has been considered in the context of Web service composition, e.g.\ in the Business Process Execution Language (BPEL)~\cite{BPEL4WS03} where role binding is supported via ``partner links''. A partner link is a variable that holds a reference to a service endpoint (i.e. a concrete address). This variable can be modified anytime during the execution of a process instance. This approach assumes that the whole process is orchestrated by a single entity and that this entity unilaterally decides which actor (i.e.\ endpoint) should be bound or re-bound to a role (i.e.\ a partner link). 
The same assumption is made in BPEL4People \cite{BPEL4People05} (which extends BPEL to support human actors), in  Pautasso et al.~\cite{Pautasso05} and Lu et al.~\cite{LU2009403}. These approaches are not applicable in settings where the binding of actors to roles is not determined by a single actor.
%proposes a task-activity-based access control (TBAC) model to manage dynamic permissions for performing tasks in a process. 
%, which studies how bindings can be established during design, compilation, deployment, process instantiation, or service invocation in service orchestrations
%\cite{LU2009403} propose a task-activity-based access control (TBAC) model that combines activities and dynamic permissions related to tasks in a business process. The implementation is compatible with BPEL. Nevertheless, all these approaches are for centralized orchestrations, and we are dealing with decentralized processes. 

%% I don't see how to fit this here and it's a short paper anyway...
%%Similarly, in \cite{DBLP:PourmirzaDG14} the parties can be replaced at runtime,  but always requires a central party serving as an intermediary.

%In a decentralized setting, \cite{Barros05} describes patterns for service integration, implementing the notion of routed interactions in BPEL using ``partner links''. 

Other work studies the problem of dynamic role binding in settings where the process is not orchestrated by a single actor. \cite{Robinson06}~extracts dynamic authorization policies from service choreographies. These policies are enforced locally by each party, but they rely on a centralized authority to specify the role bindings.
%based on the concepts role, task, and membership.
BPEL4Chor~\cite{4279612} allows an actor to bind other actors to the roles it has control over. But each role is controlled by a single actor -- in other words, collaborative role binding is not supported, e.g.\ this approach does not support the scenario where both the buyer and seller must agree on the actor who plays the role of carrier. Also, BPEL4Chor does not support role re-binding.
%, which have some similarity to the proposal presented in this paper. But additionally, 
%In contrast to our approach, releasing a participant and endorsing operations at runtime are not possible. 
%Although choreographies are usually recommended to represent the dynamics of decentralized process collaborations \cite{Mendling18}, they have not been adopted by industry to a large extent yet. Some of the challenges derived from choreographies can be found in \cite{FdhilaIRR15}. 
In~\cite{WAINER2007365,Bussard2009}, dynamic role bindings in decentralized processes are captured via a delegation and revocation scheme. This approach supports un-binding (revocation) but does not support collaborative binding (each actor decides on the roles it has control over).
%Finally, to the best of our knowledge, prior work on collaborative processes or blockchain applications in general did not consider endorsements of role bindings, thereby permitting a single party to assign roles; this is insufficient in true decentralized settings.
 %the trust cannot be delegated; then it is still an issue. 
%Finally, to the best of our knowledge, no previous work considers consensus and endorsements as a trust mechanism to handle the dynamic binding of actors.

In summary, none of the above studies has addressed the problem of dynamic role binding and un-binding in decentralized processes, where multiple actors must collaboratively agree on each role binding and un-binding decision.

%% file: model.tex
% !TEX root = ../paper.tex

\vspace{-0.7em} 
\section{Role Binding Model}
\label{sect:model}
\vspace{-0.4em} 

%\subsection{Overview}
%\label{sect:motivation}

\begin{figure*}[hbtp]
\vspace*{-12mm}
 	\hfil
 	\subfloat [Root process: Order-to-Cash]{
 		\label{subfig:sub1}
 		\includegraphics[scale=0.41]{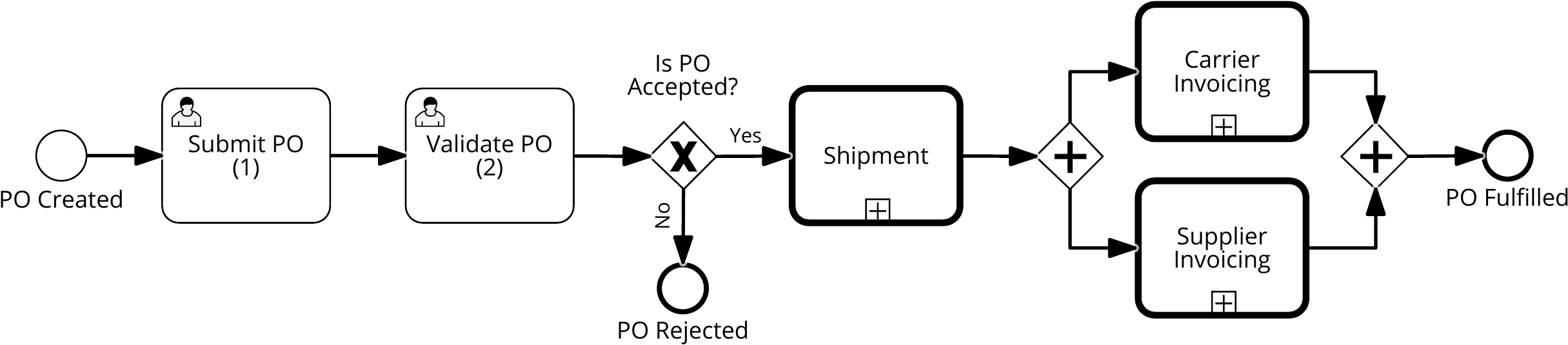}
 	}
 
    \vspace*{-6mm}
 
 	\subfloat [Sub-process: Shipment]{
 		\label{subfig:sub2}
 		\includegraphics[scale=0.41]{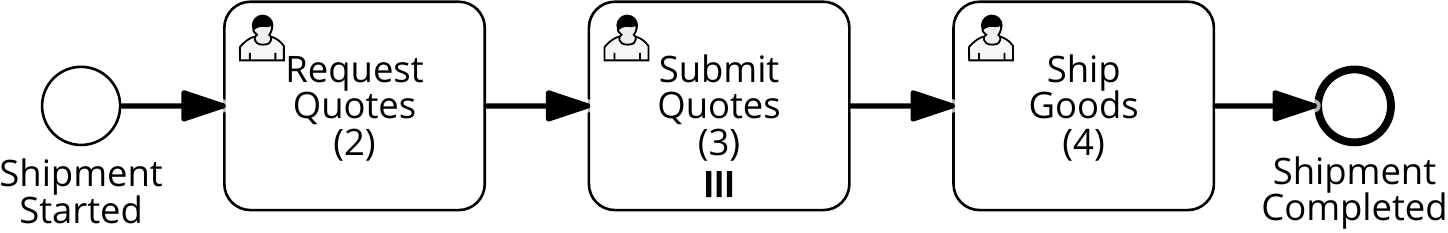}
 	}
 	\subfloat [Sub-process: Invoicing] {
 		
 		\label{subfig:sub3}
 		\includegraphics[scale=0.41]{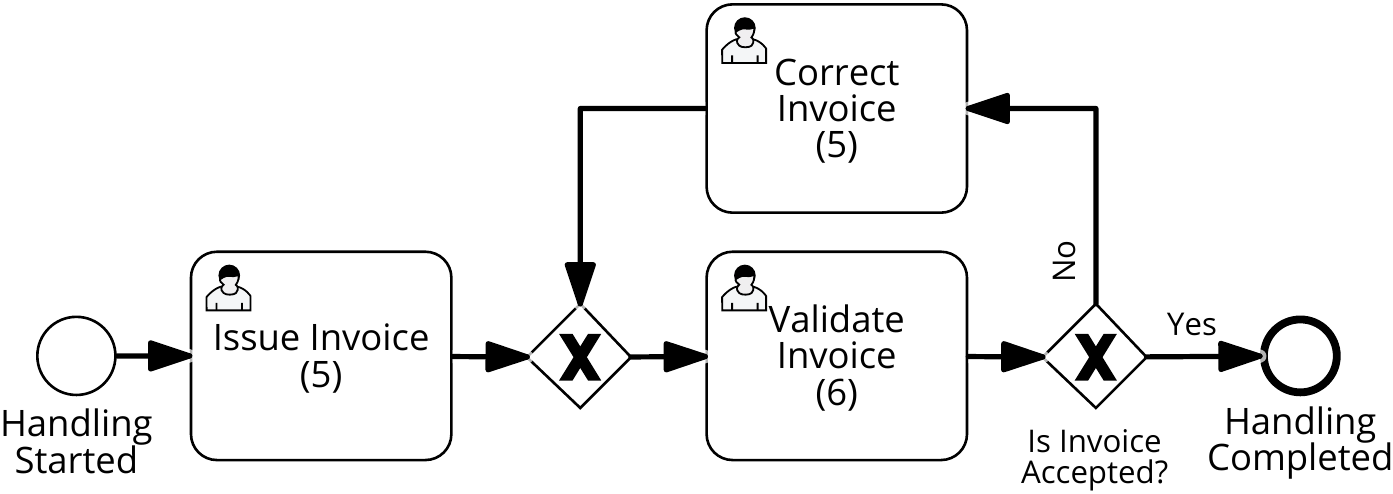}
 	}
 	\caption{\label{fig:running} Running example: (\ref{subfig:sub1}) An \textit{Order-to-cash} process linked, via call activities, to two reusable sub-processes; (\ref{subfig:sub2}) \textit{Shipment} and  (\ref{subfig:sub3}) \textit{Invoicing}.}
 	\vspace*{-6mm}
 \end{figure*}

The starting point of the proposed approach is a (collaborative) business process model where each task is associated with a role.
For a given process instance (herein called a \emph{case}), each role may be assigned to at most one actor. An actor has an identity (e.g.\ a blockchain account) and may represent a user,  a group, an organization,  a system or a device (cf. Fig.~\ref{fig:concept-relationship}).
As a running example, Fig. \ref{fig:running} shows a BPMN model of an order-to-cash process. There are six roles represented by numbers below each task label: (1) Customer, (2) Supplier, (3) CarrierCandidate, (4) Carrier, (5) Invoicer and (6) Invoicee.
%The process execution is governed by the semantics defined in the BPMN standard \cite{bpmnspec}. 
Initially, a customer submits a purchase order (PO) to a supplier. If the PO is rejected the process terminates. Otherwise, the execution continues with the {\sc Shipment} sub-process, where a supplier requests quotes from multiple carrier candidates (cf.~the multi-instance task). Once the shipment completes, two parallel paths are taken to handle the payments. These payments are encapsulated in sub-process {\sc Invoicing}.  This sub-process is called twice: for the supplier's invoice and for the carrier's invoice.
% in the root process {\sc Order to Cash}

\enlargethispage{0.5\baselineskip}
The act of assigning an actor to a role within a case is called \emph{binding}.
When a role is not assigned to an actor within a case, we say that the role is unbound.
The binding of an actor to a role can be performed anytime during the execution of a case.
Actors can also be unbound from a role -- an operation called \emph{release}.
%Role bindings are subject to scopes, like call subprocesses or multi-instance tasks.

A task is performed by the actor bound to the task's role. If a task is enabled when its associated role is unbound, the task waits until the role is bound.% to an actor.

%The basis of our proposal is that the actors of a process are appointed at runtime, i.e., during the process instance execution. An actor is a single resource, human or not, which participates in the instance of a process, e.g., by performing tasks. To allow for flexibility of actor assignment, we operate on the generic abstraction of roles, e.g., Customer, Supplier, or Carrier in the example below, and each (non-automatic) task is assigned exactly one role.

Actors may \emph{nominate} themselves or other actors to play a role in a case, or they may request to release themselves or other actors from a role.
%to accomplish different responsibilities during the process execution. 
%In case of under-performance or misbehavior, it may be necessary to release an actor from the duties assigned.
Given the lack of trust, the nomination/release of an actor to/from a role may require the endorsement of actors playing other roles. 
If an actor is nominated to play a role in a case, this nomination only leads to the role's binding if the required endorsements are granted.
A \emph{binding policy} associated to a process model determines which role(s) are allowed to nominate an actor to a role, to request the release an actor from a role, and to endorse a nomination or a release request. 

\vspace{-1mm}
\subsection{Binding Policy Specification Language}

A policy consists of a set of roles and a set of statements restricting how an actor may be nominated/released to/from a role. A statement is formed by a nominator, a nominee, and optionally a binding and/or an endorsement constraint. The nominator is a role that nominates/releases the actors of another role, namely the nominee. A binding constraint is a boolean expression stipulating that the nominee must be bound to an actor who is also bound to some other role(s). 
An endorsement constraint is a boolean expression that determines which roles need to endorse a nomination/release request. A role may be associated with the {\sf case-creator}, implying that the role is bound upon case creation and does not need a nomination or endorsement. A policy statement applies by default to the root process, but it can be scoped to a sub-process call activity. 
Fig. \ref{fig:grammar} shows an extract of the grammar of the policy language in Backus Naur form (BNF).\footnote{Some details (e.g. path expressions to refer to nested subprocesses) are omitted for space reasons and can be found at \url{http://git.io/caterpillar}.}

 \begin{figure}[htp]
 	\vspace*{-6mm}\centering\scriptsize
\begin{tcolorbox}
	\vspace*{-5mm}
\begin{grammar}
<statement> ::= [Under <subprocess> `,' ] <role> <binding\_expr> [ <endorse\_expr> \{ `,' 
 	\ind[3.8] <endorse\_expr> \} ] `;'  
 	\indalt[3.8] `case-creator' <role>`;'
 		
 	<binding\_expr> ::= (`nominates' | `releases') <role> [<binding\_constraint>]
 	
 	<binding\_constraint> ::= (`in' | `not in') <set\_expr>
 		
 	<endorse\_expr> ::= `endorsed-by' <set\_expr>
 		
 	<set\_exp> ::= <role>
 	\indalt[2.8] <role> (`and' | `or') <set\_expr>
   \indalt[2.8] `(' <set\_exp> `)'
\end{grammar}
\vspace*{-5mm}
\end{tcolorbox}
    \vspace*{-4mm}
 	\caption{\label{fig:grammar} BNF grammar describing the basic statement syntax of a binding policy. }
 	\vspace*{-5mm}
\end{figure}

%In the grammar, the nominator is the {\tt role} starting the {\tt statement} definition. The nominee is the {\tt role} in the rule {\tt binding\_expr}, which settles whether the statement is to nominate or release. The endorsement and binding constraints are respectively produced by the rules with similar names. 

%Besides, a special case of nomination statement defines which role would create the instances of related processes. 
%Note that Fig \ref{fig:grammar} only shows the key rules to describe statements, e.g., it is omitted that a role is described by its identifier.
 
 %The policy language was designed to be declarative and simple, so it is readable by stakeholders without the need of any specialized knowledge. 
Listing \ref{lst:state} shows a policy for the model in Fig. \ref{fig:running}. The policy states that the case creator is automatically bound to the {\tt Customer} role. The Customer nominates the {\tt Supplier} (no endorsement needed here). The {\tt Supplier}, in turn, nominates the {\tt Candidate} (i.e.\ the carrier candidate) and the {\tt Carrier}. The {\tt Carrier} must be among the actors bound to the {\tt Candidate} role (cf.\ binding constraint ``Carrier in Candidate'').
Note that {\tt Candidate} is a role associated to a multi-instance task ({\tt Submit Quotes}), implying that multiple actors may be bound to this role. 
%The binding constraint  stipulates that the actor to be bound to role Carrier must be among these actors.
The {\tt Customer} must endorse the nomination of the {\tt Carrier}. Under the Carrier Invoicing call activity, the {\tt Invoicer} is nominated by the {\tt Carrier} with endorsement from the {\tt Supplier} and {\tt Customer}, and reciprocally for the {\tt Invoicee}. Meanwhile, under the Supplier Invoicing activity,  the {\tt Supplier} nominates the {\tt Invoicer} with {\tt Customer} endorsement, and reciprocally for the {\tt Invoicee}.
 
%\vspace*{-3mm}
\begin{lstlisting}[language=json,label=lst:state,caption=Binding Policy to control the execution of the processes modeled in Fig. \ref{fig:running}.]
{	Customer is case-creator;
	Customer nominates Supplier;
	Under Shipment, Supplier nominates Candidate;
	Under Shipment, Supplier nominates Carrier in Candidate endorsed-by Customer;
	Under Carrier Invoicing, Carrier nominates Invoicer endorsed-by Supplier and Customer;
	Under Carrier Invoicing, Customer nominates Invoicee endorsed-by Carrier;
	Under Supplier Invoicing, Supplier nominates Invoicer endorsed-by Customer;
	Under Supplier Invoicing, Supplier nominates Invoicee endorsed-by Customer; 
}
\end{lstlisting}

%When relating a process model with a binding policy, the runtime operations provide a flexible mechanism to control the process execution. 
%For example, consider the model in Fig. \ref{fig:running} and the policy Listing \ref{lst:state}. 
This example illustrates the possibilities offered by the policy language to deal with lack of trust.
%The {\tt Supplier} can take advantage of the multi-instance task, to allow multiple candidates to submit quotes via different case scopes. 
For example, dishonest suppliers could try to derive benefits by not selecting the best carrier candidate but their preferred one. However, the customer would be able to reject such nominations. Also, the policy prevents the supplier from selecting a carrier that has not been a carrier candidate before.
%, forcing the {\tt Supplier} to select a new {\tt Carrier}.

The policy language also allows us to state that the set of actors who endorse a nomination request must fulfill a boolean expression. For instance, the above policy requires that the Invoicer of the carrier services must be endorsed by both the buyer and the supplier. This scenario is relevant in the context of international trade, where both buyers and suppliers need to ensure that they do not deal with black-listed entities or entities in countries banned from trading. The boolean expressions in the endorsement constraint may contain arbitrary combinations of conjunctions and disjunctions. They may not however contain negation, e.g.\ it is not possible to state that the nomination is approved if a given actor refuses to endorse it. Such scenarios are not applicable in this setting.

%Similarly, the {\tt Invoicer} can be nominated by {\tt Carrier} or {\tt Supplier}. Such {\tt Invoicer} would perform two tasks in the sub-process {\sc Invoice Handling} to be instantiated twice in the hierarchy, and producing two case scopes. The intuition is that each case scope will be handled by a different actor, {\tt Supplier} or {\tt Carrier}. But one of them could try to take advantage and nominate an actor in both scopes. However, the roles that can be affected with such nomination were added as endorsers. Thus they would reject a harmful nomination.         

%For the sake of reusability, the binding policy is defined independently of the process model. Thus, 
%, or on the contrary, a task may be enabled in the control flow, but no actor is appointed to perform it yet. Accordingly, a set of runtime rules must be satisfied when relating a policy with a process model.

%I think I already explained this above.
%The nomination/release of an actor may occur at any time during the execution of a case. For example, the actor who will perform the last task of the process may be nominated at the beginning of the case, or at any other point in time. It may even happen that the last task is enabled and the nomination has not happened yet.

\subsection{Runtime Binding Operations}

The role binding model relies on three operations. The {\tt nominate} operation allows an actor to request that another actor (or itself) be bound to a role within a process instance (herein called a \emph{case}). Inversely, a {\tt release} operation allows an actor to request that another actor (or itself) be unbound from a role. The {\tt vote} operation allows an actor to accept/reject a nomination or release request. 

\begin{figure}[hbt]
 	\vspace*{-5mm}
 	\centering
 	\includegraphics[width=0.95\textwidth]{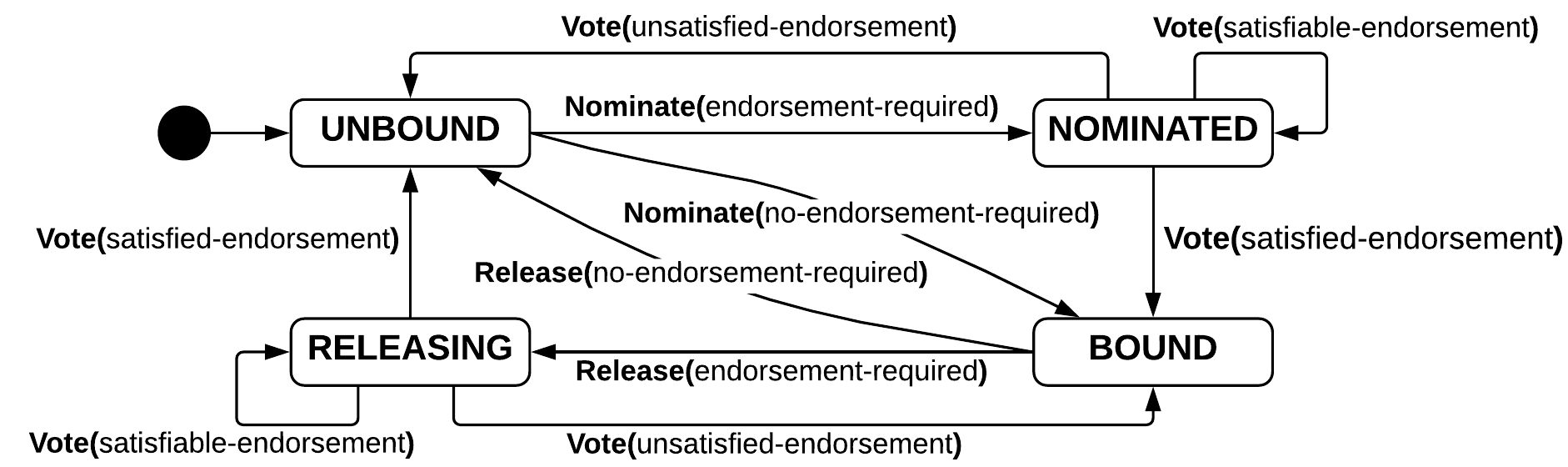}
 	\vspace*{-2mm}
 	\caption{\label{fig:binding-states} Lifecycle of a role within a case. }
 	\vspace*{-5mm}
 \end{figure}

These operations trigger transitions in the \emph{role lifecycle} depicted in Fig.~\ref{fig:binding-states}. Within a case, a role is initially {\sc unbound}. After a {\tt nominate} operation, the role changes to {\sc nominated} if it requires to be endorsed, otherwise is considered {\sc bound}. A role in {\sc nominated} state, can transition to the {\sc bound} state after a {\tt vote} operation where the endorser accepts the nomination if, as a result of it, the endorsement constraint of this role is satisfied. On the contrary, a {\tt vote} operation where the endorser rejects the nomination and by doing so makes the role's endorsement constraint unsatisfiable, triggers a transition to the {\sc unbound} state. If after a {\tt vote} operation, the endorsement constraint remains satisfiable, then the role remains in the {\sc nominated} state. Symmetrically, a role can transit from {\sc bound} to {\sc unbound} as a result of a {\tt release} operation, via a {\sc releasing} state, which is specular to the {\sc nominated} state. If the endorsement constraint associated to a \emph{release} request becomes unsatisfiable, the role goes back to the {\sc bound} state, and if it becomes satisfied, the role moves to the {\sc unbound} state.

%% file: consistency.tex
\section{Policy Consistency Verification}
\label{sect:policy}

%  As stated in section \ref{sect:model} a binding policy is described by a set of statements. An statement consists of \textit{nominator}, \textit{nominee}, \textit{binding} and \textit{endorsement} constraints. A statement restricts how a \textit{nominee} can be nominated or released at runtime. Then, a policy is consistent if every statement guarantees that the nomination or release of an actor as \textit{nominee} can reach state BOUND/UNBOUND at runtime.
  
%  The \textit{nominate} and \textit{release} operations implicitly establishes a precedence relation between the roles involved, i.e. the nominate operation requires a nominator BOUNDED to an actor in advance. Only those roles described as case-creators or allowed to self-nominate by the binding policy are exempt of appointing a nominator in advance. Besides, a \textit{nominee} can be nominated only if the \textit{binding constraint} is satisfied, and an eventual endorse requires fulfilling the \textit{endorsement constraint}. Both constraints are described by set expressions comprising roles and two boolean operators, \textit{and} / \textit{or}. Accordingly, the roles in such expressions must be BOUND before the \textit{nominee} can reach such state, thus they must be considered in the precedence relation.

Nomination and release statements in a policy implicitly induce precedence dependencies in the binding of roles. A statement {\tt R1 nominates R2 endorsed-by R3} implies that for R2 to be bound, R1 and R3 must be bound before.
%Only those roles bound to {\tt case-creator} are exempt from such dependencies. 
Circular and unresolvable dependencies induced in this way may lead to deadlocks. 
%Besides, a \textit{nominee} can be nominated only if the \textit{binding constraint} is satisfied, and an eventual endorse requires fulfilling the \textit{endorsement constraint}. Both constraints are described by set expressions comprising roles and two boolean operators, \textit{and} / \textit{or}. Accordingly, the roles in such expressions must be {\sc Bound} before the \textit{nominee} can reach such state, thus they must be considered in the precedence relation.
Accordingly, we define a notion of  policy consistency as follows. A policy is \emph{consistent} if, starting from the state where only the roles associated with case-creator are {\sc Bound} and after executing any allowed sequence of nomination, release and endorse operations, we always reach a state where all roles will reach the {\sc Bound} state via some (other) sequence of nomination, release and endorse operations. 

To verify policy consistency, we define a mapping from a policy to a Petri net~\cite{Murata89}, herein called a \emph{nomination net}. Given the nomination net of a policy, we map the problem of checking policy consistency to a problem of reachability analysis over Petri nets.
%To formalize this notion and to enable the verification of policy concisten, we construct a net, called , which encodes the statements of a binding policy and over which the aforementioned reachability analysis can be performed. In the following, we introduce the method, starting with some definitions.
%A Petri net is a directed graph with two types of nodes, places and transitions, hence bipartite.
% Its arcs connect either a place to a transition or vice versa. 
%Graphically, places are represented as circles and transitions as boxes. The marking of a net associates a nonnegative number of tokens to each place. Each token is represented as a dot inside the corresponding place. A transition is considered enabled if each of its input places contains at least one token. When an enabled transition ``fires'', one token is removed from each input place and one put into each output place.
Algorithm \ref{algo:nom:net} maps a policy to a nomination net. For the sake of conciseness, this algorithm focuses on nomination statements, leaving aside release statements. The mapping of release statements follows a similar structure. For the same reason, the algorithm leaves aside binding constraints.

\begin{algorithm}[t!]
	% \textbf{Input: } BP(R, S): Binding Policy \\
	% \textbf{Output: } NP: Nomination Net  
\algnewcommand{\LeftComment}[1]{\Statex \textcolor{blue!40}{\(\triangleright\) #1}}
\algblockx{Let}{EndLet}[1]{\textbf{let} #1 \textbf{in}}

\begin{algorithmic}[1]
\Function{ConstructNominationNet}{R, BP}
\State RNets $\gets \emptyset$
\LeftComment {Step 1: Build a Petri net for each role}
\For{{\bf each} role $r \in$ R}
	\State RNets $\gets$ RNets 
 $
 	\bigcup \left\{\left( r \mapsto \left< 
 	\begin{array}{lll}
		\{u_r, n_r, b_r\} &\phantom{.}& \triangleright \ P_r\\
		\{nm_r, en_r\} && \triangleright \ T_r \\	
		\{(u_r, nm_r), (nm_r, n_r), (n_r, en_r), (en_r, b_r)\} && \triangleright \ F_r \\	
	\end{array}
 	\right> \right) \right\}
 $ 
\EndFor
\LeftComment {Step 2: Merge all role nets to form the nomination net}
\Let {NNet $= \left< P, T, F, M_0 \right>$}
	\State $P\ \gets\ \bigcup_{r \in \text{R}}\mathcal{P}(\text{RNets}[r])$
	\State $T\ \gets\ \bigcup_{r \in \text{R}}\mathcal{T}(\text{RNets}[r])$
	\State $F\ \gets\ \bigcup_{r \in \text{R}}\mathcal{F}(\text{RNets}[r])$
	\State $M_0\ \gets\ \emptyset$
\EndLet

\LeftComment {Step 3: Wire up operation \sc{nominate}}
\For {{\bf each} $\left<r_{nr}, r_{ne}, \_ \right> \in$ BP}
	%, NNet $= \left< P, T, F, M_0 \right>$}
	\State {\bf select} $b_{nr} \in \mathcal{P}(\text{RNets}[nr])$
	\State {\bf select} $nm_{r_{ne}} \in \mathcal{T}(\text{RNets}[ne])$
	\State $\mathcal{F}(\text{NNet}) \gets \mathcal{F}(\text{NNet}) \cup \{ (b_{r_{nr}}, nm_{r_{ne}}), (nm_{r_{ne}}, b_{r_{nr}})\}$
\EndFor

\LeftComment {Step 4: Wire up operation \sc{endorse}}
\For {{\bf each} $\left<r_{nr}, r_{ne}, eex \right> \in$ BP {\bf such that} $eex \neq \bot$}
	%, NNet $= \left< P, T, F, M_0 \right>$}
	\State {$\mathcal{P}(\text{NNet}) \gets \mathcal{P}(\text{NNet}) \cup \{ disj_{r_{ne}}, eex_{r_{ne}} \} $}
  \State {$\mathcal{F}(\text{NNet}) \gets \mathcal{F}(\text{NNet}) \cup \{ (nm_{r_{ne}},  disj_{r_{ne}}), (eex_{r_{ne}}, en_{r_{ne}}) \}$}
	
	\For {{\bf each} conj $\in eex$}
		\State {$\mathcal{T}(\text{NNet}) \gets \mathcal{T}(\text{NNet}) \cup \{ eex_{conj} \} $}
\State $\mathcal{F}(\text{NNet}) \gets \mathcal{F}(\text{NNet}) \bigcup\limits_{r \in conj \land b_r \in \mathcal{P}(\text{RNets}[r])} \left\{\begin{array}{l} 
(b_r, eex_{conj}), (eex_{conj}, b_r),\\ 
(disj_{r_{ne}}, eex_{conj})
\end{array} \right\}$
	\EndFor
\EndFor

\LeftComment {Step 5: Update NNet's initial marking}
\Let{$r_{cc} \in R$: $r_{cc}$ be case creator}
	\State Ps $\gets \{ u_r\ |\ r \in R \setminus \{r_{cc}\} \land u_r \in \mathcal{P}(\text{NNet}[r]) \} \cup \{ b_{r_{cc}}\ |\  b_{r_{cc}} \in \mathcal{P}(\text{NNet}[r_{cc}]) \}$

	\State $M_0(\text{NNet})(p) = \left\{ \begin{array}{lll} 1 &\phantom{MMMMMM}& \text{if }p \in Ps\\ 0 & & \text{Otherwise}\\ \end{array} \right.$
\EndLet
\State \Return NNet
\EndFunction
\end{algorithmic}
\caption{Construction of the Nomination Net for a given Binding Policy}
\label{algo:nom:net}
\end{algorithm}

\begin{figure}[t]
	\begin{minipage}[t]{0.45\textwidth}
	% \tiny
		\begin{verbatim}
{ A is case-creator;
  A nominates B;
  A nominates C;
  C nominates D, endorsed-by A and B; 
}
		\end{verbatim}
		\vspace{-5.5mm}
		\caption{\label{binding:policy}Sample binding policy}
	\end{minipage}
	\begin{minipage}[t]{0.46\textwidth}
		$$
		\begin{matrix}
			R = \{A, B, C, D\}\\
			BP = \{\left<A, B, \bot \right>, \left<A, C, \bot \right>, \left<C, D, A\wedge B \right>\}
		\end{matrix}
		$$
		\vspace{-5mm}
		\caption{\label{symbolic:rep}Symbolic representation of the binding policy in Fig.~\ref{binding:policy}}
	\end{minipage}
	\includegraphics[scale=.5]{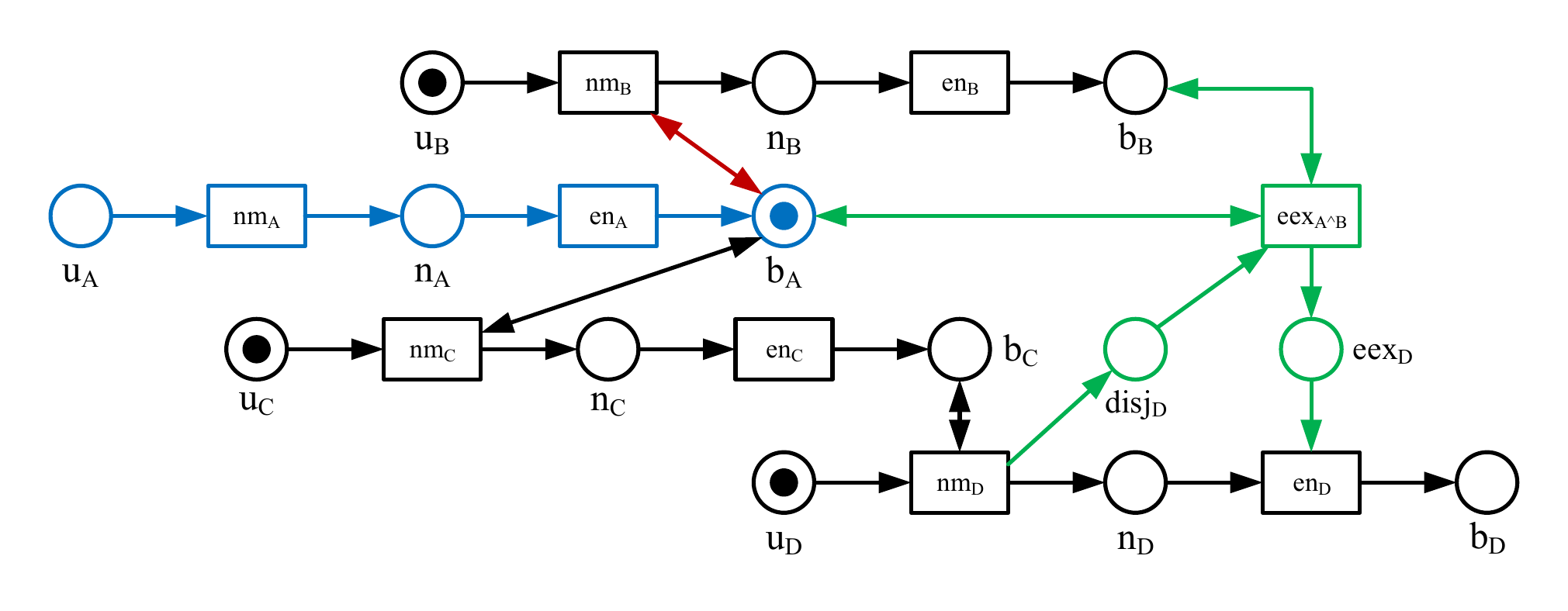}
	\vspace{-3mm}
	\caption{\label{fig:nom:net1} Nomination net for binding policy in Fig.~\ref{binding:policy}}
	\vspace{-8mm}
\end{figure}

%The method for building the nomination net for a given binding policy is outlined in Algorithm \ref{algo:nom:net}.
To illustrate the algorithm, we consider the binding policy in Fig.~\ref{binding:policy}. The algorithm takes as input a symbolic representation of a policy consisting of a set of roles and a set of tuples of the form (nominator, nominee, endorsement-constraint), with $\bot$ denoting an empty constraint.
For example, the symbolic representation of the policy in Fig.~\ref{binding:policy} is given in Fig.~\ref{symbolic:rep}.
Given this input, the algorithm will produce as output the nomination net in Fig.~\ref{fig:nom:net1}.

%In the specification of Algorithm~\ref{algo:nom:net} we chose to use a symbolic representation (see Fig.~\ref{symbolic:rep}) for the internal representation, which forms the inputs passed to function {\sc ConstructNominationNet} in Algorithm~\ref{algo:nom:net}.
The algorithm proceeds as follows. After initializing variable \verb|RNets| in line 2, the algorithm builds a Petri net for each node in lines 3-4 (Step 1).
Let us consider that we are building the Petri net for role $A$, which is shown in color blue in 
Fig.~\ref{fig:nom:net1}. In line 4, the algorithm creates such a Petri net with three places, namely
$u_A$, $n_A$ and $b_A$, which represent the states of the role's lifecycle UNBOUND, NOMINATED and BOUND, respectively.
Similarly, two transitions are added to the Petri net, namely $nm_A$ and $en_A$, representing the
operations 'nominate' and 'endorse'. Finally, four arcs added to complete the Petri net, by connecting the places and transitions. 
The Petri nets for all the other nodes are created in a similar way. Every Petri net thus created
is added to \verb|RNets| that serves as a map that associates a role to its corresponding Petri net.

In lines 5-9 (Step 2), all the role (Petri) nets are merged to form the initial nomination net, which is held in
variable \verb|NNet|. This is done by taking the union of the elements in the role nets. 
Also, the initial marking is set to the empty set.

In lines 11-14 (Step 3), the algorithm adds double-headed arcs to the Petri net to synchronize the transition that represents
the nomination of roles. To illustrate the idea of nomination, consider the double-headed arc connecting the place $b_A$
and the transition $nm_B$ in Fig.~\ref{fig:nom:net1}, highlighted in red. 
Simply put, role $A$ will be able to nominate role $B$ 
when role $B$ is UNBOUND and role $A$ is BOUND ($b_A$ must hold a token). The firing of transition $nm_B$, that is "nominate B",
will change the state of role $B$ from UNBOUND to NOMINATED. The double-headed arc will keep a token in
$b_A$ after the nomination of role $B$.

The encoding of endorsement conditions is handled in lines 15-20 (Step 4).
Without loss of generality, we assume that the endorsement conditions are expressed in disjunctive normal form, meaning
that there is only one disjunction that relates several conjunctions. We consider two additional cases: (1) no endorsement condition
is specified (represented by $\bot$), meaning that no endorsement is required, and (2) only one conjunction is specified.
To illustrate this step of the construction of the nomination net, consider the binding policy:
$$
\text{D nominates E, endorsed-by (A and B) or (B and C);}
$$

\begin{wrapfigure}{r}{.33\textwidth}
	\vspace*{-6mm}
	\includegraphics[width=\linewidth]{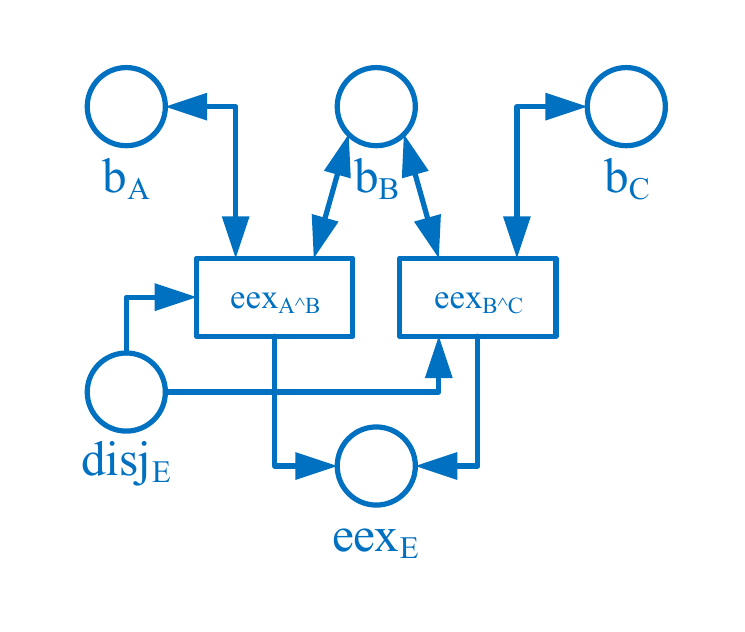}
	\vspace*{-10mm}
	\caption{\label{fig:endorsement:net}Net encoding condition $(A\land B)\lor(B\land C)$}
	\vspace*{-5mm}
\end{wrapfigure}

The Petri net in Fig.~\ref{fig:endorsement:net} encodes the endorsement
condition in the above policy: $(A\land B) \lor (B\land C)$. 
The latter is bound to variable $eex$ in line 15.
In line 16, the algorithm adds two new places: $disj_E$ which encodes the disjunction, 
and $eex_E$, which collects the outcome of the endorsement (i.e. it holds a token
when one of the endorsement conditions is met).
In line 17, these are connected to the transitions of the role: from the nomination $nm_E$ to $disj_E$, and from the outcome $eex_E$ to the endorsement $en_E$ (not shown in Fig.~\ref{fig:endorsement:net}).
Then, in line 18, the algorithm iterates over each one of the conjunctions. In line 19,
a new transition, representing the underlying conjunction is added to the net, and the corresponding arc in line 20. For instance,
the net in Fig.~\ref{fig:endorsement:net} has transition $eex_{A\land B}$ representing
conjunction $A\land B$, and $eex_{B\land C}$ representing $B\land C$. 
Only $eex_{A\land B}$ or $eex_{B\land C}$ will be able to consume the token held by $disj_E$, which prevents the generation of an arbitrary number of tokens in \verb|NNet|. 
$disj_E$ receives a token when $nm_E$ fires, i.e., when $D$ nominates $E$.
The disjunction expressed in this way means that role $E$
can be endorsed if at least one of the conjunctions holds true, which corresponds to
the firing of one of the transitions $eex_{A\land B}$ and $eex_{B\land C}$.
Returning to the example in Figures~\ref{binding:policy}-\ref{fig:nom:net1}, we observe that role $D$ is endorsed if and only if both roles $A$ and 
$B$ are BOUND. The subnet implementing the endorsement condition is shown in green in
Fig.~\ref{fig:nom:net1}.

Finally, lines 21-23 set the initial marking for the nomination net. Briefly, line 21 will
add a token to the place representing the state UNBOUND of every single role, except for the
``case creator''. In the latter case, we add a token to the place representing the state BOUND.

\begin{figure}[h]
\vspace{-2mm}
\begin{minipage}[t]{0.32\textwidth}
	\begin{verbatim}
{
 J is case-creator;
 J nominates K, endorsed-by L;
 J nominates L, endorsed-by K;
}
	\end{verbatim}
	\vspace{-4mm}
\end{minipage}
~~~	~	
\begin{minipage}[t]{0.65\textwidth}
  \vspace{-8mm}
	\includegraphics[scale=.5]{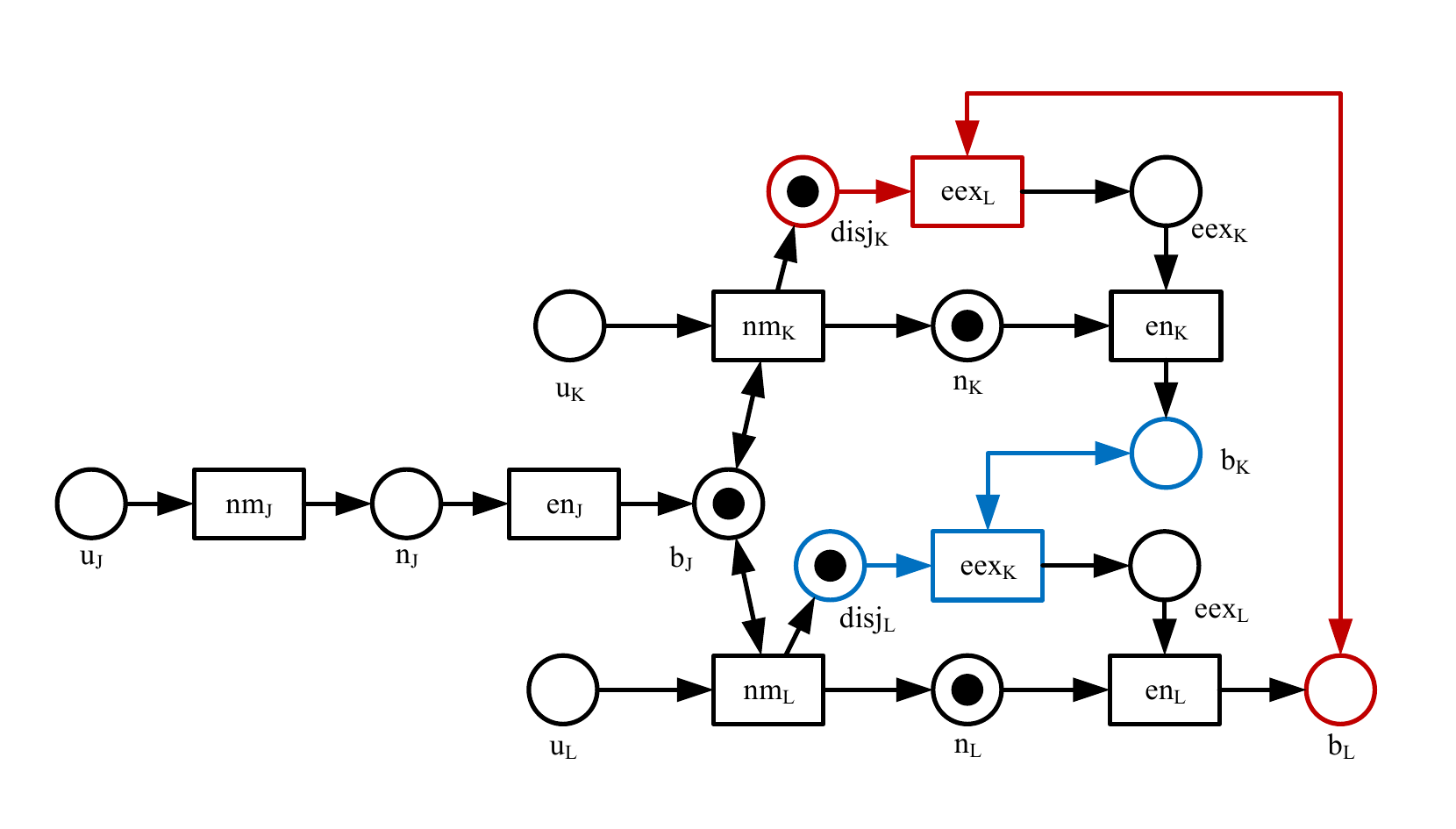}
  \vspace{-8mm}
\end{minipage}
\caption{\label{fig:deadlocked} Binding policy with circular dependency and its nomination net}
  \vspace{-6mm}
\end{figure}

To verify policy consistency, we use reachability analysis to check if the marking where all roles are bound is always reachable starting from the initial marking where only the roles associated to {\tt case-creator} are bound. In other words, there is no deadlock preventing a role from being bound.
Fig.~\ref{fig:deadlocked} shows a binding policy with a circular dependency, leading to a deadlock in the corresponding nomination net. Fig.~\ref{fig:deadlocked} shows the marking where the deadlock occurs. Both roles $K$ and $L$ have been
nominated by role $J$. Hence, $disj_K$ has a token, but transition $eex_L$ cannot
fire until $b_L$ has also a token. In order for $b_L$ to have a token, however, transition $eex_K$ needs to
fire because it requires $b_K$ to have a token. 

%% file: implementation.tex
\section{Implementation and Evaluation}
\label{sect:implementation}

To demonstrate the proposal's feasibility, we developed a compiler that takes as input a policy specification and produces Solidity smart contracts to enforce the policy. This policy compiler is designed to be used in conjunction with the {\sc Caterpillar} BPMN-to-Solidity compiler~\cite{Lopez-PintadoGDWP18}. The smart contracts generated by the policy compiler manage the association between roles and actors (represented as blockchain accounts), while the smart contracts generated by the BPMN-to-Solidity compiler enforce the control-flow constraints in the process model. When a task is enabled, the \emph{worklist handler} smart contract of {\sc Caterpillar}, checks if the corresponding role is bound to an actor within the current case, and ensures that only this actor can execute the task. 
The source code of {\sc Caterpillar}, including the binding policy compiler and the examples used in this paper, are available at \url{http://git.io/caterpillar}.
%The prototype allows, via REST interactions, the validation of binding policies that are compiled later into smart contracts in Solidity. Besides, it supports to perform the runtime operations, i.e., nomination, vote, release, as well as executing process models restricted by our access control approach. 
Below we discuss the generation of smart contracts and evaluate the costs generated by these contracts.% in Ethereum.
%generated by the policy compiler and an empirical evaluation of the costs generated by these smart contracts on the Ethereum platform.

\subsection{Compiling Binding Policies into Smart Contracts} \label{sect:comp}
 
Given a process model and a policy specification, the policy compiler generates a smart contract (named {\sc BindingPolicy}) to encode the policy and a smart contract ({\sc TaskRoleMap}) to encode the task-role relations in the process model. 
The {\sc BindingPolicy} contract encodes the logic of who can nominate and release each role and the binding and endorsement constraints for each role.
A third contract ({\sc BindingAccessControl}) implements the runtime operations sketched in Section~\ref{sect:model}.  {\sc BindingPolicy} and {\sc TaskRoleMap} are singleton contracts -- only one instance of each of them is created since these contracts only maintain schema-level data. Meanwhile, the {\sc BindingAccessControl} contract is instantiated once per case. 
The {\sc BindingAccessControl} contract instance of a given case maintains the state of each role, as per the lifecycle in Fig.~\ref{fig:binding-states}. When a nomination, release, or vote operation is invoked, the {\sc BindingAccessControl} contract invokes the {\sc BindingPolicy} contract. The latter checks if this operation is allowed in the current state and computes the new state.

%These contracts are generated from the policy specification and the process model and hence they do not write/read data from the blockchain storage which can be costly. 
%. Besides, this contract would be independent of the policy, i.e., it does not store any data from the policy, but it keeps references to the {\sc BindingPolicy} and {\sc TaskRoleMap} contracts to query such information.

\begin{figure}[htp]
  	\vspace*{-4mm}
 	\centering
 	\includegraphics[width=\textwidth]{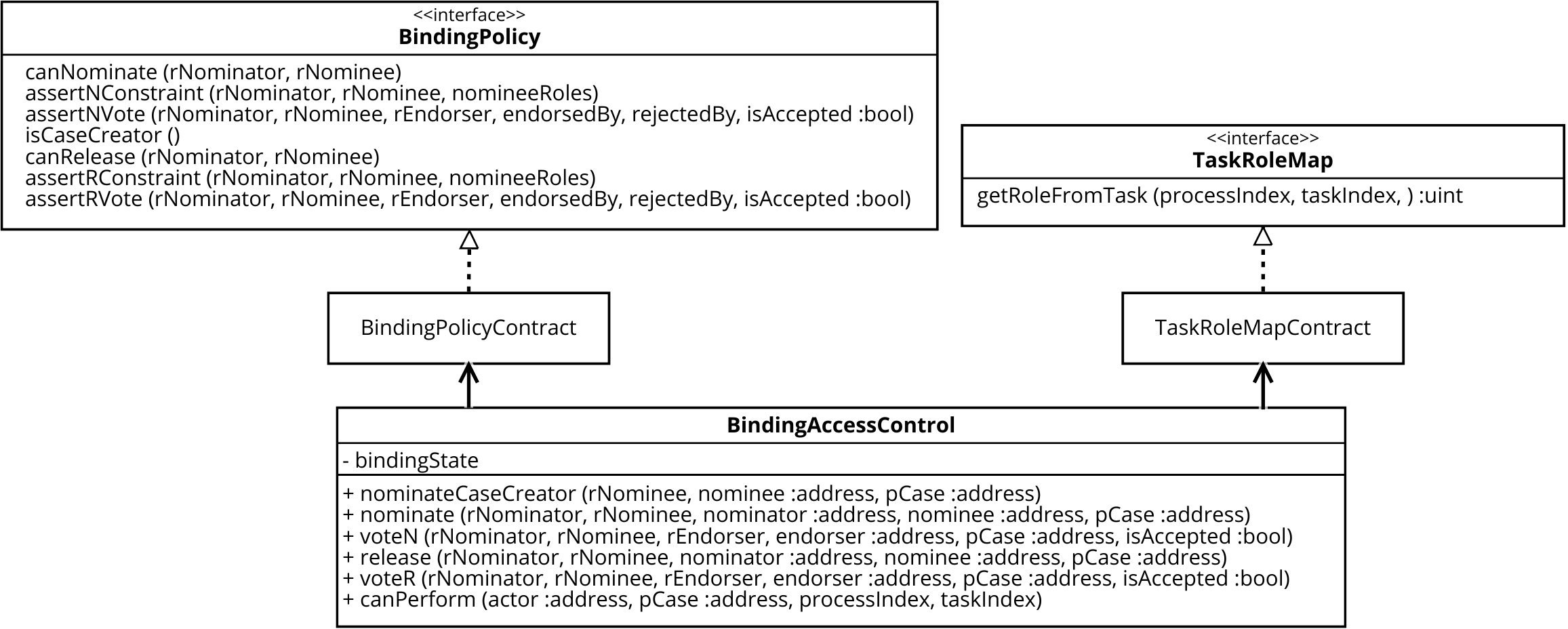}
 	\vspace*{-2mm}
 	\caption{\label{fig:class-diagram} Class diagram of the smart contracts derived from the policies. }
 	\vspace*{-5mm}
 \end{figure}

The class diagram in Fig \ref{fig:class-diagram} captures the functionality of the generated smart contracts. Input parameters with no type specification are by default {\tt uint}. 
As stated above, contract {\sc BindingAccessControl} implements the  runtime operations for nomination, release and voting. Since this contract does not encode anything about a particular policy, it is not generated by the policy compiler, but instead it is hard-coded and deployed once on the target Ethereum blockchain. 
This contract maintains the state of the role bindings for a given case in a variable called {\sc bindingState}. Given that the cost of a smart contract depends on the amount of data it maintains, we encode the {\sc bindingState} using bitmaps. 
%Specifically, we assign a sequential number (starting from 1) to each role. If a role is {\sc Unbound}, we represent it with a zero. 
Similarly, the endorsement constraints are represented as bit arrays. 
Specifically, we first put these constraints in disjunctive normal form, e.g., {\tt (A and B and ...) or (D and ...)}. Then we implement each conjunction set as a bit array and encode it as a 256-bits unsigned integer -- the default word size in Ethereum.
 
Contract {\sc TaskRoleMap} is generated from the process model. This contract is straightforward (it maps tasks to roles), so we do not discuss it further.
%: Function {\tt getRoleFromTask} has one conditional statement per task encoding the role associated to this task. 
% (the same for all policy specifications).  

The policy specification is compiled into the {\sc BindingPolicy} contract.
Below we discuss how the role binding functions are generated (functions {\tt canNominate}, {\tt assertNConstraint} and {\tt assertNVote}).  The generation of the release functions ({\tt canRelease}, {\tt assertRConstraint} and {\tt asserRVote}) is done in a similar way. 

To generate function {\tt canNominate}, for each distinct nominator in the policy a conditional and bit array, namely {\tt nMask}, is created with one bit per role such that the presence of a nominee is represented with a \emph{one} and the absence with a \emph{zero}. For example, a nominator with index 3 and {\tt nMask = 6} is translated into:
 
 \begin{tcolorbox}
 	\vspace*{-4mm}
\begin{lstlisting}[language=Solidity]
function canNominate(uint rNominator, uint rNominee) returns(bool) {
		...
		if (rNominator == 3)
			return 6 & (1 << rNominee) != 0;
		...
}
\end{lstlisting}
 	\vspace*{-5mm}
 \end{tcolorbox}
 
Function {\tt assertNConstraint} verifies if the roles held by a nominee do not contradict the binding constraint. Thus, a conditional instruction per nomination statement including a binding constraint is added. A statement is identified by the union of nominator and nominee, i.e., {\tt (1 << rNominator) | (1 << rNominee)}. Variable {\tt nomineeRoles} is the bit array encoding the nominee's current roles. Given a constraint of the form {\tt (A and B) or (C) or ..}, the constraint is fulfilled if at least one of the conjunction sets is fully included in {\tt nomineeRoles}. This is encoded as follows:
 
\begin{tcolorbox}
\vspace*{-4mm}
\begin{lstlisting}[language=Solidity]
if ((1 << rNominator) | (1 << rNominee))
	return nomineeRoles & ((1 << A) | (1 << B)) == ((1 << A) | (1 << B)) 
 		   || nomineeRoles & (1 << C) == (1 << C) || ...;
\end{lstlisting}
\vspace*{-4mm}
\end{tcolorbox}

Function {\tt assertNVote} checks if an endorser can vote for a nomination and determines the state after this vote. Given the input parameters {\tt endorsedBy} and {\tt rejectedBy}, which are bit arrays encoding the roles that already accepted and rejected the nomination, this function determines the resulting state as follows:

 \vspace*{-1mm}
 \begin{enumerate}
 	\item {\sc Bound} if all the roles in at least a conjunction set, namely {\tt CS}, endorsed the nomination, i.e., {\tt (endorsedBy | endorserRole) \&  CS == CS},
 	\item {\sc Unbound} if in each conjunction set contains at least one role rejected the nomination, i.e., for each {\tt CS}, {\tt (rejectedBy | endorserRole) \& CS != 0},
 	\item {\sc Nominated} if none of the conditions 1. and 2. are fulfilled yet, i.e., there is at least a conjunction set with no rejections and with roles pending to vote.
 \end{enumerate}
   
 \subsection{Experimental Setup}
 
We conducted an evaluation to answer the following question: How does the cost (in gas/ether) of enforcing a binding policy increase depending on the size and complexity of the policy statements?\footnote{In Ethereum, gas is linearly related to throughput, see Section~\ref{ssect:blockchain-bg}. So by answering this question we also indirectly answer the related throughput question.}  We decompose this question into three: (Q1) How do the costs of deploying the generated smart contracts vary with the size of the policy? (Q2) How do the costs of executing the runtime operations vary with the size of the policy? (3) How does the combined cost of enforcing a process model and a binding policy varies with the size of the model?

%generation of smart contracts from a policy is deterministic. Explicitly, the 
It follows from Section \ref{sect:comp} that the costs depend on the number of roles to nominate and the number of conjunction sets in the binding/endorsement constraints. Thus, we designed the following experiments: (E1) We varied the number of nomination statements in a policy from 1 to 40, without any binding or endorsement constraints. (E2) We fixed the number of statements to 40, selected one statement, and gradually increased the size of its conjunction set from 1 to 40. (E3) We fixed the number of statements to 40, and gradually added a binding constraint with one conjunction set to each of the 40 statements. (E4,E5) The experiments E3 and E4 were repeated for the endorsement constraint (instead of the binding constraint). (E6) We generated a policy with 40 roles such that each statement includes a binding constraint stipulating that the nominated actor must belong to the role in the previous statement and that the nomination must be endorsed by all actors nominated in previous statements. (E7) Starting from a BPMN model with only one task, we iteratively expanded it by one task at a time (up to 40) and assigned each task to a different role. In this latter experiment, once a role was bound to an actor, we checked that the corresponding task could be performed. 
%We run 40 iterations for each experiment. 
% In all the cases, each role described by the policies was nominated, and the endorsements were provided when required. 
Note that the evaluation focuses on nomination statements, but the release statements are symmetric. %Thus the costs are proportional to those of the nomination statements.
 
 We implemented a replayer in Java that generates the policies, triggers their compilation and deployment, and executes the runtime operations via {\sc Caterpillar}'s REST API. For each transaction included in the blockchain, Caterpillar sends some meta-data that includes block number, consumed gas, transaction hash which is collected and assessed by the replayer. For the experimentation we run a \textit{Node.js} based Ethereum client named \textit{ganache-cli}\footnote{\url{https://github.com/trufflesuite/ganache-cli}} which is widely used to simulate a full client for developing and testing purposes on Ethereum.
 
\subsection{Experimental Results}
  
 \begin{wrapfigure}{r}{0.5\textwidth}
  	\vspace*{-9mm}
  	\centering
  	\includegraphics[width=0.5\columnwidth]{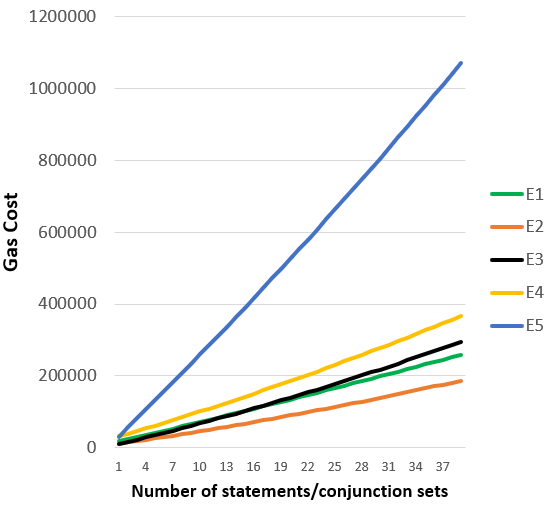}
  	\caption{Growth of deployment costs with size of binding policy.}
  	\label{fig:deployment-cost}
  	\vspace*{-6mm}
 \end{wrapfigure}
 
Deployment costs for experiments E1-E5 are plotted in Fig. \ref{fig:deployment-cost}. 
It can be seen that deployment costs increase quasi-linearly with the size and complexity of the policy. The simplest contract (with a single role bound to case-creator) costs 154,167 gas. 
%Then, we computed how the costs grew relative to the base contract for experiments E1-E5. 
As expected, the most pronounced growth in cost occurs for endorsement constraints (E4-E5) as they produce more instructions during code generation. We observe an increase of around  $16.0-19.0\%$ when adding a new endorsement constraint and $5.0-6.5\%$ when adding a conjunction set to a constraint. Experiments E2-E3 show that adding a binding constraint increases cost by $4.0-5.7\%$, while adding a conjunction to a constraint adds  $2.4-3.5\%$ overhead. E1 shows that adding one unrestricted statement to nominate a role adds $4.0-4.5\%$ overhead.   

\begin{wraptable}{r}{0.52\textwidth}
%\vspace*{-8mm}
\scriptsize
\begin{tabular}{|c|l|r|r|r|r|r|}
\hline
&& \multicolumn{1}{c|}{\textbf{E1}} & \multicolumn{1}{c|}{\textbf{E2}} & \multicolumn{1}{c|}{\textbf{E3}} & \multicolumn{1}{c|}{\textbf{E4}} & \multicolumn{1}{c|}{\textbf{E5}}\\
\hline
\parbox[t]{2mm}{\multirow{3}{*}{\rotatebox[origin=c]{90}{Nom.}}} & Min. & 151,586 & 112,476 & 111,407 & 132,417 & 131,493\\
& Max. & 152,638 & 152,790 & 113,447 & 152,746 & 153,800 \\
& Ave. &151,948 & 151,270 & 112,277 & 151,738 & 142,660 \\
\hline
\hline
\parbox[t]{2mm}{\multirow{3}{*}{\rotatebox[origin=c]{90}{Vot.}}} & Min. & - & - & - & 76,845 & 77,184\\
& Max. & - & - & - & 78,136 & 78,184 \\
& Ave. & - & - & - & 77,463 & 77,541 \\
\hline
\end{tabular}
\caption{Cost of nominations and votes}
\label{tbl:results}
\vspace*{-7mm}
\end{wraptable}

%Although the variations were not linear, the differences and costs were considerably smaller compared to the deployment costs.
For the runtime operations, we observed that costs vary depending on the number and the order of statements and conjunction sets in the constraints. The cost to nominate a role is higher when the corresponding policy statement is at the end of the policy. Similar behavior was observed for binding and endorsement constraints. This is because in Ethereum, the gas depends on the number of bytecode instructions executed. Hence, in a function with {\tt if-else-if} instructions, the cost increases with the number of evaluated conditions. Table \ref{tbl:results} shows the min, max, and average costs to perform the nominate and vote operations in experiments E1-E5. Note that  voting is less costly than nominating and nomination costs are lower when restricted by binding constraints compared to endorsement constraints.

The combined cost of executing a process model with an associated policy (experiment E6) has several components.
First,   
%Overall, the cost to handle the process execution with policy is given by (1) 
the contract {\sc BindingAccessControl} must be deployed at a fixed cost of 1,340,098 gas, entailing a transaction fee of 0.0067 Ether (ETH)\footnote{Gas price: 5 Gwei, average from \url{https://ethgasstation.info} on 30/11/2018.}. %, approx. US \$0.38. %\iw{TBH, I wouldn't bother with US\$ numbers.}
Next, the contracts generated from the policy must be deployed, with gas ranging from 154,167 (simplest) to 1,803,898 gas (largest policy), corresponding to 0.0007 ETH to 0.0090 ETH. % (US \$0.04) ... (US \$0.51). 
The smart contracts derived from the policies are deployed once and then reused, while the contracts handling the process execution are deployed for each case. Thus the policy deployment costs are amortized as more cases are executed. At runtime, roles have to be bound to actors. The costs of executing one nominate operation ranged from 111,407 (0.0005 ETH) to 168,270 (0.0008 ETH), while the {\tt vote} operations cost between 76,845 (0.0003 ETH) and 78,184 (0.0003 ETH). 
%(0.0003 ETH, US \$0.03) to 168270 (0.0004 ETH, US \$0.05), while the {\tt vote} operations were between 76845 (0.0002 ETH, US \$0.02) and 78184 (0.0002 ETH, US \$0.02). 
Finally, when an actor performs a task, function {\tt canPerform } is called to check if the actor is bound to the task's role. This function invokes the {\sc TaskRoleMap} smart contract to retrieve the task-role relation. We observed a linear growth in the deployment cost of this contract as the number of tasks increased, from 129,539 gas (0.0006 ETH) to 241,114 (0.0012 ETH). The cost of function {\tt canPerform} also grew linearly from 31,693 (0.0001 ETH) to 33,066 (0.0001 ETH).

%with the contract {\sc TaskRoleMap} 
%(0.0003 ETH, US \$0.04) to 241114 (0.0006 ETH, US \$0.07). Similarly, the cost of the function {\tt canPerform} grows linearly with the contract {\sc ProcessPolicy} from 31693 gas (0.00008 ETH, US \$0.009) to 33066 (0.00008 ETH, US \$0.009).
 
 % The experiments comprised a set of more than 200 policies with different levels of complexity, which intentionally verifies the cases that increase the execution costs. Accordingly, results offer an approximate estimation of the costs to expect when instantiating/executing policies containing among 1-40 roles.

%% file: conclusion.tex
% !TEX root = ../paper.tex

\section{Conclusion}
\label{sect:conclusion}

Motivated by the possibilities opened by blockchain-based collaborative process execution, this paper presented a role binding model and a binding policy language that support collaborative binding and unbinding of actors to roles at runtime. %The approach provides a flexible and dynamic access control mechanism to execute collaborative processes on blockchain systems. 
The proposal includes a method to verify the consistency of policies defined in the proposed language and  % to avoid deadlocks on the process execution. Besides, 
an approach to compile the policies into smart contracts. The proposal has been implemented on the {\sc Caterpillar} blockchain-based collaborative process execution tool.
%, including respective backend and front-end extensions. 
%To that end, the extension of Caterpillar includes a compiler of binding policies into Solidity, and a panel to deploy the smart contracts and execute the binding operations. 
%Indeed, the execution of processes in Caterpillar is controlled by the binding policies. 
%In the experiments, 
We evaluated the costs (and therefore throughput) to deploy and execute smart contracts generated from binding policy statements, on the Ethereum platform.
 %Although 
The evaluation shows that the deployment and runtime policy enforcement costs grow linearly with the number of roles and the complexity of the constraints.
%which we plan to optimise in future work. 
 %On the other hand, the endorsements serve to restrict, for example, that no actor can be appointed to perform a critical task without the approval of other participants that can be affected by its execution. However, the endorsements cannot entirely prevent scenarios when an actor leaves during the process execution affecting other participants. For example, an anonymous customer submits a purchase order, and never shows up when the goods are produced. This issue can be solved off-chain, for example by forcing the customer to pay a security deposit before proceeding with the endorsement. However, extending the policy language to prevent such scenarios via smart contracts, is another direction of future work.
%
We acknowledge that the evaluation is limited in scope (only one business process and up to 40 roles) and focuses on evaluating cost. An avenue for future work is to further validate the approach via more thorough experiments and case studies. 

While the proposed approach has been designed with the goal of supporting collaborative process execution on blockchain, its field of possible applications is wider. Another future work avenue is to study the applicability of this approach to other blockchain applications where dynamic role binding may be required, e.g.\ in crowdsourcing and computer-supported collaborative work scenarios.

%% file: paper.bbl
\begin{thebibliography}{10}
\providecommand{\url}[1]{{#1}}
\providecommand{\urlprefix}{URL }
\expandafter\ifx\csname urlstyle\endcsname\relax
  \providecommand{\doi}[1]{DOI~\discretionary{}{}{}#1}\else
  \providecommand{\doi}{DOI~\discretionary{}{}{}\begingroup
  \urlstyle{rm}\Url}\fi

\bibitem{BPEL4WS03}
Andrews, T., et~al.: {BPEL4WS, Business Process Execution Language for Web
  Services Version 1.1}.
\newblock IBM (2003)

\bibitem{Bussard2009}
Bussard, L., Nano, A., Pinsdorf, U.: Delegation of access rights in
  multi-domain service compositions.
\newblock Identity in the Information Society \textbf{2}(2), 137--154 (2009)

\bibitem{4279612}
Decker, G., Kopp, O., Leymann, F., Weske, M.: {BPEL4Chor}: Extending {BPEL} for
  modeling choreographies.
\newblock In: IEEE ICWS 2007, pp. 296--303 (2007)

\bibitem{DBLP:FrantzN16}
Frantz, C., Nowostawski, M.: From institutions to code: Towards automated
  generation of smart contracts.
\newblock In: {IEEE} {FAS*W} 2016, pp. 210--215 (2016)

\bibitem{BPEL4People05}
Kloppmann, M., et~al.: {WS-BPEL} extension for people - {BPEL4People}.
\newblock Joint white paper, {IBM} and {SAP}  (2005)

\bibitem{Lopez-PintadoGDWP18}
L{\'{o}}pez{-}Pintado, O., Garc{\'{\i}}a{-}Ba{\~{n}}uelos, L., Dumas, M.,
  Weber, I., Ponomarev, A.: {Caterpillar:} {A} business process execution
  engine on the ethereum blockchain.
\newblock CoRR \textbf{abs/1808.03517} (2018)

\bibitem{LU2009403}
Lu, Y., Zhang, L., Sun, J.: Task-activity based access control for process
  collaboration environments.
\newblock Computers in Industry \textbf{60}(6), 403--415 (2009)

\bibitem{Mendling18}
Mendling, J., et~al.: Blockchains for business process management - challenges
  and opportunities.
\newblock {ACM} Trans. Management Inf. Syst. \textbf{9}(1), 4:1--4:16 (2018)

\bibitem{Murata89}
Murata, T.: Petri nets: Properties, analysis and applications.
\newblock Proceedings of the IEEE \textbf{77}(4), 541--580 (1989)

\bibitem{Pautasso05}
Pautasso, C., Alonso, G.: Flexible binding for reusable composition of web
  services.
\newblock In: Software Composition, pp. 151--166 (2005)

\bibitem{Prybila0HW17}
Prybila, C., Schulte, S., Hochreiner, C., Weber, I.: Runtime verification for
  business processes utilizing the {Bitcoin} blockchain.
\newblock Fut. Gen. Comp. Syst. \textbf{46}, 36--50 (2017)

\bibitem{Robinson06}
Robinson, P., Kerschbaum, F., Schaad, A.: From business process choreography to
  authorization policies.
\newblock In: Data and Applications Security, pp. 297--309 (2006)

\bibitem{RussellAHE05}
Russell, N., van~der Aalst, W.M.P., ter Hofstede, A.H.M., Edmond, D.: Workflow
  resource patterns: Identification, representation and tool support.
\newblock In: CAiSE (2005)

\bibitem{TranLW18}
Tran, A., Lu, Q., Weber, I.: Lorikeet: {A} model-driven engineering tool for
  blockchain-based business process execution and asset management.
\newblock In: Demo Track at {BPM} 2018, pp. 56--60 (2018)

\bibitem{WAINER2007365}
Wainer, J., Kumar, A., Barthelmess, P.: {DW-RBAC}: A formal security model of
  delegation and revocation in workflow systems.
\newblock Inf. Syst. \textbf{32}(3), 365 -- 384 (2007)

\bibitem{2019-Blockchain-Book}
Xu, X., Weber, I., Staples, M.: Architecture for blockchain applications.
\newblock Springer (2019)

\end{thebibliography}
